\documentclass{pasj00}

\begin{document}
\SetRunningHead{Kajisawa et al.}{DRG Number Counts in MOIRCS Deep Survey}
\Received{}
\Accepted{}

\title{MOIRCS Deep Survey. I: DRG Number Counts}

%
\author{Masaru \textsc{Kajisawa},\altaffilmark{1} 
Masahiro \textsc{Konishi},\altaffilmark{2,3} 
Ryuji \textsc{Suzuki},\altaffilmark{3} 
Chihiro \textsc{Tokoku},\altaffilmark{3} \\
Yuka \textsc{Katsuno Uchimoto},\altaffilmark{4} 
Tomohiro \textsc{Yoshikawa},\altaffilmark{2,3} 
Masayuki \textsc{Akiyama},\altaffilmark{3} 
Takashi \textsc{Ichikawa},\altaffilmark{2} \\
Masami \textsc{Ouchi},\altaffilmark{5,6}
Koji \textsc{Omata},\altaffilmark{3} 
Ichi \textsc{Tanaka},\altaffilmark{3} 
Tetsuo \textsc{Nishimura},\altaffilmark{3} 
Toru \textsc{Yamada},\altaffilmark{3}}
\altaffiltext{1}{National Astronomical Observatory of Japan, Mitaka, Tokyo
181--8588, Japan}
\email{kajisawa@optik.mtk.nao.ac.jp}
\altaffiltext{2}{Astronomical Institute, Tohoku University, Aramaki,
Aoba, Sendai 980-8578, Japan}
\altaffiltext{3}{Subaru Telescope, National Astronomical Observatory
of Japan,\\ 
650 North Aohoku Place, Hilo, HI 96720, USA}
\altaffiltext{4}{Institute of Astronomy, University of Tokyo, Mitaka, Tokyo
181-0015, Japan}
\altaffiltext{5}{Space Telescope Science Institute, 3700 San Martin Drive,
Baltimore, MD 21218, USA}
\altaffiltext{6}{Hubble Fellow}

%


\maketitle

\begin{abstract}
We use very deep near-infrared imaging data  
taken with Multi-Object InfraRed Camera and Spectrograph (MOIRCS) on
the Subaru Telescope to investigate the number  
counts of Distant Red Galaxies (DRGs). We have observed a 4$\times$7 
arcmin$^{2}$ field 
in the Great Observatories Origins Deep Survey North (GOODS-N), 
and our data reach J=24.6 and K=23.2
(5$\sigma$, Vega magnitude). 
The surface density of DRGs selected by $J-K>2.3$ is 
2.35$\pm$0.31 arcmin$^{-2}$ at $K<22$ and 3.54$\pm$0.38
 arcmin$^{-2}$ at $K<23$, respectively.
These values are consistent with those in the GOODS-South and FIRES. 
Our deep and wide data suggest that the number counts of DRGs
 turn over at $K\sim22$, and the surface density of the faint DRGs
with $K>22$ is smaller than that expected from the number counts at
the brighter magnitude. 
The result indicates that while there are many bright  
galaxies at $2<z<4$ 
with the relatively old stellar population and/or heavy dust extinction, 
the number of the faint galaxies with the similar red color is relatively
small. 
Different behaviors of the number counts of the DRGs and  bluer 
galaxies with $2<z_{\rm phot}<4$ 
at $K>22$ suggest that the mass-dependent color distribution, 
where most of low-mass galaxies are blue  
while more massive galaxies tend to have redder colors,  
had already been established at that epoch.
\end{abstract}

\section{Introduction}
\label{sec:intro}

The Distant Red Galaxies (DRGs) are selected in the near-infrared
(NIR) wavelength by the simple criterion $J-K>2.3$ in order to sample
high-redshift ($z\gtrsim2$) galaxies with significant fraction 
of evolved stars such as 
normal galaxies seen in the present universe, which are often missed 
by the ``drop-out'' selection technique for the Lyman break galaxies
 (\cite{fra03}). 
These red $J-K$ colors are produced by Balmer/4000\AA-break 
at $2\lesssim z\lesssim4$ (the breaks enter between $J$ and
$K$-bands) and/or heavy dust extinction at similar redshift.
In fact,  several DRGs are confirmed to be at $2<z<4$ by spectroscopy 
(\cite{van03}, \cite{red05}), and their photometric redshift also 
 lies in the range of $2\lesssim z\lesssim4$ (\cite{for04}).

Several studies found that 
many DRGs have high star formation rates
(\cite{van04}, \cite{kun05}, \cite{red05}), 
while some of them seem to have little star formation activity and
evolve passively (\cite{lab05}, \cite{pop06}).
Analyses of the spectral energy distributions (SEDs) suggest
that DRGs are more massive (M$_{*}\gtrsim10^{11}$M$_{\odot}$) and older
($\sim$1-3 Gyr old) than the UV-bright LBGs at similar redshifts
(\cite{for04}, \cite{iwa05}, \cite{lab05}, \cite{pop06}, \cite{kri06}). 
Recently, the strong angular clustering of these galaxies has also 
been reported 
(\cite{gra06}, \cite{fou06}, \cite{qua06}).

However, most of these previous studies have focused on the bright
(massive) part of DRGs. It is important to sample fainter DRGs and 
investigate their properties in order to understand the formation and
evolution of more ``normal'' galaxies. Galaxies with the 
luminosity of L$^{*}$ or sub-L$^{*}$ at $z\sim3$ 
 would have K-band magnitude of $K\sim$22-23.  
Such very deep NIR data have been limited to relatively  
small areas so far (e.g., \cite{mai01}, \cite{lab03}, \cite{min05}).

In this paper, we use very deep and wide 
NIR data taken with MOIRCS on the Subaru Telescope in order to 
investigate the number counts of DRGs down to $K=23$. 
The wide field of view of MOIRCS ($4\times7$ arcmin$^{2}$) and the
large collecting area of the telescope allow us to analyze DRGs to the
faint magnitude with high statistical accuracy. 
 
The Vega-referred magnitude system is used throughout the paper.

\section{Observation and Data Analysis}
\label{sec:obs}
We performed $JHK_s$-bands imaging observations of a part of the Great
Observatories Origins Deep Survey North (GOODS-N) field 
with Multi-Object InfraRed Camera and Spectrograph
 (MOIRCS, \cite{ich06}) on the Subaru Telescope  
on 2006 April 4-9,18 and May 10-11,18-19. MOIRCS has the field of view of
about 4 $\times$ 7 arcmin$^{2}$ with 0.117 arcsec pixel scale. 
We observed one field of view of MOIRCS centered on
12$^{h}36^{m}46^{s}.62$, 
 $+62^{\circ}13'15''.6$ (J2000), which included the
original Hubble Deep Field North (HDF-N, \cite{wil96}).   
Here we use the high-quality $J$ and $K_s$-bands data sets which include  
 only the frames with the seeing size smaller than 0.5 arcsec (FWHM). 
The corresponding exposure times
are 5.0 hours in $J$-band and 7.7 hours in $K_s$-band.

The data were reduced in a standard way using the IRAF software package.
At first the self-skyflat frame was made and used for the
flat-fielding. We then performed the sky subtraction, and
co-registered and combined the data.
The details of the data processing and the data quality are described 
in Konishi et al. (in preparation).
The FWHMs of the PSFs of the combined images are 0.42 arcsec
 in $J$-band and 0.40 
arcsec in $K_s$-band, and 
the $K_s$-band image was convolved with a Gaussian kernel to match the
PSF to that in $J$-band.
FS23 and FS27 in the UKIRT faint standard stars
 were used for the flux calibrations in the  
$J$ and $K$-bands.
We excluded the edges of the images with lower sensitivities, and  
our final multi-band combined images have a slightly smaller field
of view of about 24.3 arcmin$^{2}$.

Source detection was performed in the $K_s$-band image using the
SExtractor image analysis package (\cite{ber96}). 
We adopted MAG\_AUTO from the SExtractor as the total $K$-band
magnitudes of the detected objects. For the color measurements, 
we used the fixed aperture with 0.85 arcsec diameter (2$\times$FWHM),
and the same apertures were used for $J$ and $K$-bands.

The 5$\sigma$ limiting magnitude at 0.85 arcsec diameter is about 
 $J=24.6$ and $K=23.2$, respectively. 
We estimated the 1$\sigma$ background fluctuation in each band 
 by directly measuring the sky
 fluxes with 0.85 arcsec apertures randomly placed on the images.
The estimated 
 background fluctuations were used to calculate the photometric errors 
for the detected objects. 
We performed the simulations 
using the IRAF/ARTDATA package to quantify the detection completeness in
 $K$-band. Our source detection is nearly 
complete for the point source to $K\sim23$, 
and the completeness is 90\% at $K\sim23.3$. 
We also tested sensitivity to the false detections by running
 SExtractor on the inverted $K_s$-band image. Only 15 spurious
 objects were extracted at $K<23$, while 1596 objects with $K<23$  
were detected in the normal image. 

\section{Results}

\label{sec:colsel}
\subsection{Selection of DRGs}
\begin{figure}
\begin{center}
\FigureFile(85mm,85mm){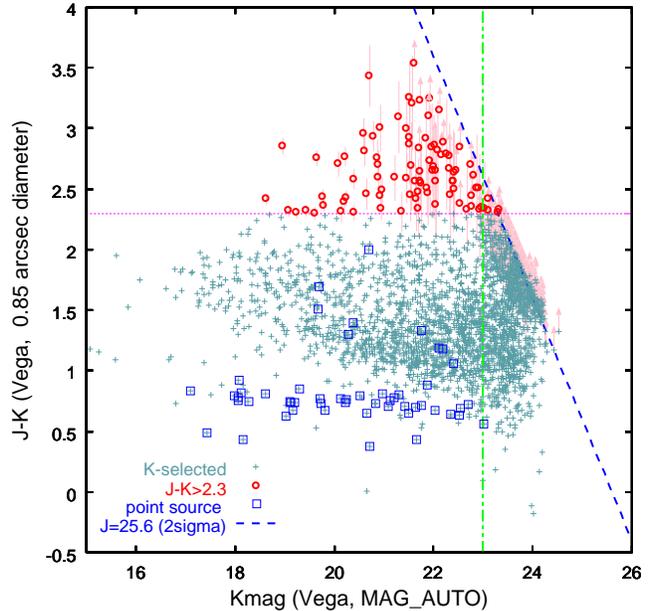}
\end{center}
\caption{Color-magnitude diagram of the objects in the 
MOIRCS deep imaging field in
the GOODS-North. 
Galaxies with $J-K>2.3$ are plotted by large circles.
Vertical dotted-dash line represents the completeness limit
at $K$-band. 
Dashed line shows the 2$\sigma$ limit at $J$-band.
For objects with $J$-band fluxes lower than the limit, 
the lower limits of $J-K$ color are plotted by symbols with arrows.
Those objects do not necessarily lie on the dashed line because 
 the aperture sizes for the $J-K$ color and the $K$-band magnitude 
 are different (0.85 arcsec diameter and the Kron aperture,
respectively).
Squares indicates the objects which  
are identified as point sources in the publicly available GOODS
HST/ACS $z$-band data.
\label{fig:jk}}
\end{figure}

Figure \ref{fig:jk} shows $J-K$ vs $K$ colour-magnitude diagram of the 
objects in the 
MOIRCS deep imaging field. We have selected DRGs by the criterion of
$J-K>2.3$. These DRGs are plotted as 
the large circles, while the others are showed as the small crosses.
The squares show the objects which were identified 
as point sources in the publicly available
GOODS HST/ACS $z$-band data (\cite{gia04}). The sequence of these objects is 
seen at $J-K\lesssim0.9$, which is consistent with the expected locus of 
the galactic stars.
This suggests that the flux calibration of our data is good enough for the
selection of DRGs.
The 2$\sigma$ limit at $J$-band and the completeness limit at $K$-band are 
also showed (dashed line and dotted-dash line). The deep $J$-band data  
(the 2$\sigma$ limit of $J=25.6$) allows us to sample the objects with
 $J-K>2.3$ down to $K\sim23.3$. 

91 (86 at $K<23$) DRGs were detected in our field of $\sim$24.3
arcmin$^{2}$. The cumulative surface density of DRGs down to $K=$21,
22, 23 is tabulated in Table 1. 
The errors of the surface densities are based on Poissonian.
The surface density of 1.15$\pm$0.22 arcmin$^{-2}$ 
at $K<21$ is consistent with those in the HDF-South (HDF-S) 
and MS1054-03 field reported by F{\"o}rster Schreiber et 
al. (2004).

%
\subsection{Number Counts of DRGs}
\begin{figure*}
\begin{center}
\FigureFile(110mm,220mm){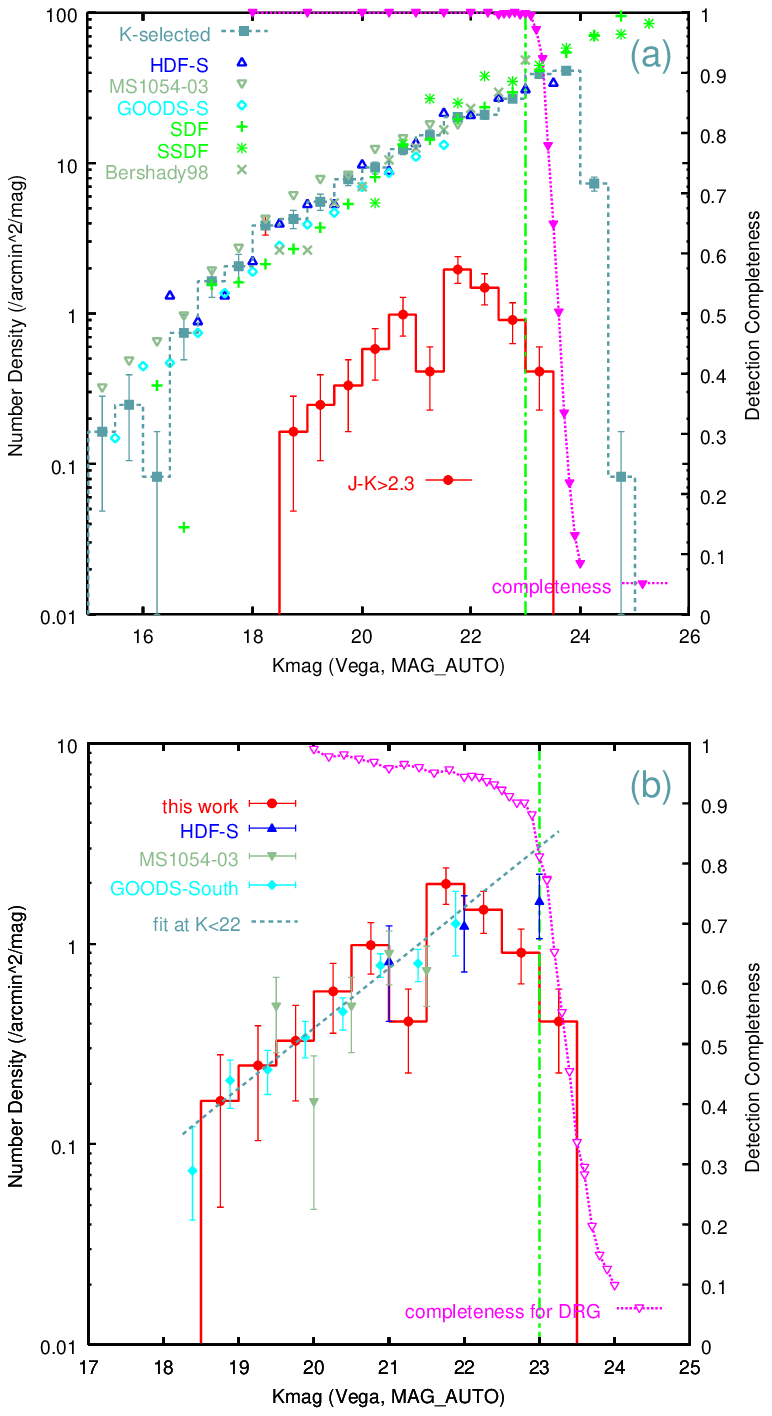}
\end{center}
\caption{ {\bf(a)} Number counts of all $K$-selected objects
(squares) and DRGs (circles) in the MOIRCS deep imaging field.
The detection completeness in $K$-band
is also shown (dotted line). The vertical dotted-dash line shows our
completeness limit. 
$K$-band number counts in other fields are also shown (HDF-South from
\cite{lab03}, MS1054-03 field from \cite{for06},
GOODS-South from \cite{gra06}, Subaru Deep Field from \cite{mai01}, 
Subaru Super Deep Field from \cite{min05}, high Galactic latitude
fields from \cite{ber98}). 
{\bf (b)} Comparison of DRG number counts. The
DRG number counts in other fields are also shown (HDF-South, 
MS1054-03, GOODS-South). The dashed line shows the linear
fit for the data points at $K<22$ in all four fields (this study, 
HDF-S, MS1054-03, GOODS-S). Errorbars are based on Poissonian (i.e.,
square roots of the observed numbers).
Dotted line shows the detection completeness for extended objects with 
FWHM$=$0.58 arcsec (see text).}

\label{fig:num}
\end{figure*}

\begin{table}
 \centering
  \caption{Number density of DRGs in the MOIRCS deep field}
  \label{tab:num}
  \begin{tabular}{@{}clcc  c @{}}
  \hline 
  \hline
& $K$ magnitude &  Number  & density {\footnotesize (arcmin$^{-2}$ mag$^{-1}$)}& \\ 
 \hline
& $<18.5$    &  0 & 0  &\\ 
& 18.5-19.0  &  2 & 0.16 $\pm$ 0.12  &\\
& 19.0-19.5  &  3 & 0.25 $\pm$ 0.14  &\\
& 19.5-20.0  &  4 & 0.33 $\pm$ 0.16  &\\
& 20.0-20.5  &  7 & 0.58 $\pm$ 0.22  &\\
& 20.5-21.0  & 12 & 1.07 $\pm$ 0.30 &\\
& 21.0-21.5  &  5 & 0.41 $\pm$ 0.18 &\\   
& 21.5-22.0  & 24 & 1.98 $\pm$ 0.40 &\\                
& 22.0-22.5  & 18 & 1.48 $\pm$ 0.35 &\\
& 22.5-23.0  & 11 & 0.91 $\pm$ 0.27 &\\
\hline
&                &        & cumulative density &\\
& $K$-band limit & Number &{\footnotesize (arcmin$^{-2}$)}& \\
\hline
& $<21.0$ & 28 & 1.15 $\pm$ 0.22 &\\
& $<22.0$ & 57 & 2.35 $\pm$ 0.31 &\\
& $<23.0$ & 86 & 3.54 $\pm$ 0.38 &\\
\hline
\end{tabular}
\end{table}

Figure \ref{fig:num} shows the number counts of the DRGs in the MOIRCS deep
imaging field. We show the 
number counts of all $K$-selected objects and the DRGs in the top panel.
The $K$-band detection completeness mentioned in the previous section is 
also plotted. 
Our data are nearly complete down to $K=23$, and the number counts in
Figure \ref{fig:num} are not corrected for the detection completeness.
We also show $K$-band number counts in other general fields from the 
literature. The number counts
of all $K$-selected objects in our data are consistent with those in
other fields down to $K=23$.

The number of the detected DRGs and their surface
density in each $K$-band magnitude bin are also showed in Table 1.
The interesting result in Figure \ref{fig:num} is that 
the number counts of DRGs turn over at $K\sim22$, while the $K$-band
number counts continue to increase to at least $K=23$, which corresponds to 
the completeness limit. 

In the bottom panel of Figure \ref{fig:num}, we compare the number counts
of DRGs in our data with those in other fields (HDF-S,
MS1054-03, GOODS-South). 
The HDF-S data by the FIRES survey (\cite{lab03}) have the smaller area 
($\sim$4.5 arcmin$^{2}$) but reach to the deeper completeness limit 
($K\sim$23.8) than our data. The areas of the GOODS-S and MS 1054-03 fields 
are comparable with or wider than this study ($\sim$130 and $\sim$25
arcmin$^{2}$, respectively), but these data reach to the shallower limit of
$K\sim$21.5-22. 

Though these independent four fields have different areas and depths, 
the number counts of DRGs in these fields agree well at $K<22$. 
At $K>22$, only the HDF-S data have the sufficient depth and can be
used for the comparison with out data. The number counts of DRGs in
the 
HDF-S do not show the decrease at $K>22$, but are still consistent with our 
results within the uncertainty (see also the next section). 

In order to demonstrate the turnover of the DRG counts or the deficit of
the faint DRGs relative to the brighter ones, we performed the linear
fitting for the number counts of DRGs at $18.5<K<22$ using the data in all
four fields. The linear fitting at $18.5<K<22$ seems to be reasonable 
because Foucaud et 
al. (2006) reported the break feature of the DRG counts lies 
at $K\sim18$ and the data over $18.5<K<22$ of all four fields were 
fitted well with the linear line.  
The result is plotted as the dashed line in Figure \ref{fig:num}.
We also tried to fit the MOIRCS data at $K<22$ with the Maximum Likelihood
method without binning the data, and confirmed the very similar 
result of the fitted slope. Thus the uncertainty due to the binning of
the data does not affect the fitting result, even if there is a
bin with the relatively low counts ($K=$21-21.5 bin).
  
When we extrapolate this linear line to $K\sim23$, the surface density 
of the faint DRGs with $22<K<23$ in our field is clearly 
deficient. The density in the $22.5<K<23$ bin is about a factor of three 
 lower than 
the extrapolation from the number counts at the brighter magnitude.
If we perform the similar linear fitting  
for the MOIRCS data over $18.5<K<23$ with the Maximum Likelihood
method, the fitted slope become 0.19 $\pm$ 0.04, which is
significantly  
flatter than the value for the fitting at $18.5<K<22$ (0.33 $\pm$
0.07).
It is seen that even in the HDF-S, 
the number of the faint DRGs is lower than the extrapolation from the 
number counts at $K<22$.
%
%


\section{Discussion}
\label{sec:overdense}
In this section, we discuss the turnover of the number counts of DRGs 
at $K\sim22$ found in the previous section. 

At first, we consider the possible spurious effects which could cause
the deficit of the faint DRGs. One possibility is that the MAG\_AUTO 
would systematically underestimate  
the total flux in the $K$-band near the detection limit (e.g.,
\cite{lab03}), and this could cause the underestimation of the
number of the faint DRGs. The number counts of DRGs, however, begin to 
decrease at $K\sim22$, which is one magnitude brighter than the {\it
completeness} limit in $K$-band, and this does not seem to be the case. 
 For example, 
Labb{\'e} et al. (2003) showed such a underestimation occurs at $K\sim24$, 
which is about 0.3 mag fainter than the completeness limit in their
FIRES data. Our simulation also indicates that the significant
underestimation of the $K$-band flux does not exist down to at least
$K=23$ (Konishi et al. in preparation). In fact, the $K$-band number
counts continues to increase to $K=23$ in Figure \ref{fig:num}.
We also confirmed that 
the size distribution of DRGs with $K\sim23$ is not significantly
different from that of the bluer objects.
Therefore there is no reason to think 
that only DRGs are less complete at $K\sim23$.

Since most DRGs in our data are resolved, the completeness problems
could be more severe than those for the point source. 
Therefore we also checked the detection completeness for extended objects. 
We fitted the $K$-band surface brightness of relatively bright DRGs
with $K<21$ with the Moffat function, and found that the average FWHM
is about 0.58 arcsec. Using this surface brightness profile (assuming 
the same profile for the fainter DRGs),  
we performed the similar simulation with that for the point source to 
quantify the detection completeness as a function of $K$-band
magnitude. 
The result is showed in the bottom panel of Figure \ref{fig:num}
(dotted line).
The completeness is more than 90\% at $K<22.5$ and is still 80\% at 
$K=23$. The detection incompleteness in $K$-band cannot explain the 
observed deficit of the faint DRGs.

Second possibility is that the larger error in $J-K$ color at the faint
magnitude affects the sampling of DRGs. Although the depth of our
$J$-band data (the 2$\sigma$ limit of $J=25.6$) is sufficient to
sample the objects with $J-K>2.3$ to $K=23$, some fraction of the
faint DRGs could be missed due to the photometric error. However, we
conclude that   
the error in $J-K$ does not cause the deficit of the faint
DRGs, since the similar effect of the contamination from the bluer
objects to the DRG sample also exists. As discussed in Foucaud et 
al. (2006),
since the overall $J-K$ distribution of objects has the peak at the 
color substantially bluer than $J-K\sim2.3$ and the objects with
$J-K>2.3$ are relatively rare (Figure \ref{fig:jk}), it can be expected that 
the number density of DRGs is rather boosted at fainter
magnitudes. Therefore these photometric errors do not seem to cause 
the observed deficit of DRGs at $K>22$.
\begin{figure*}
\begin{center}
\FigureFile(135mm,90mm){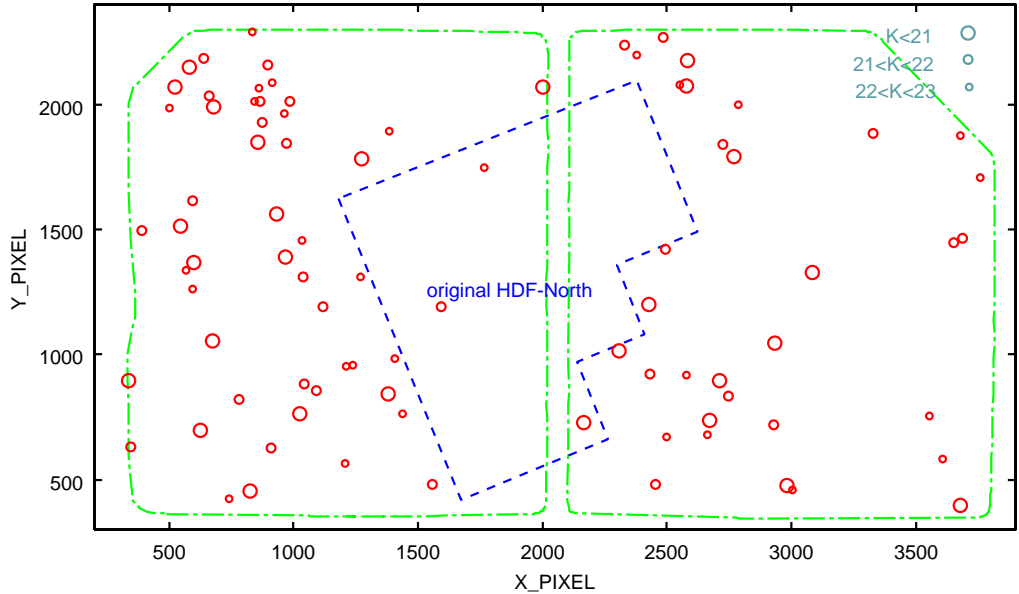}
\end{center}
\caption{Spatial distribution of the DRGs. The size of the symbols is scaled
according to apparent magnitudes in $K$-band (the bigger, the
brighter). Dashed line shows the region of the original HDF-N.
Dotted-dash lines show the field of view of the MOIRCS data.
MOIRCS achieves the field of view of $4\times7$ arcmin$^{2}$ with two 
HAWAII-2 arrays. The present version of the combined data were
separately reduced for each array, and the deep portion of the field
of view is divided into two areas. Two corners (upper left and upper
right) are vignetted at the Cassegrain focus of the telescope.
}

\label{fig:xy}
\end{figure*}

Although the number counts of DRGs in the HDF-S 
 do not show the turnover at $K\sim22$ as seen in the previous section, 
the difference of the number density of DRGs at $K>22$ between 
in the HDF-S and in our MOIRCS field can be explained by the field-to-field
variance, when the strong clustering of DRGs is taken into account 
(e.g., \cite{gra06}, \cite{qua06}). Grazian et al. (2006) pointed out the 
large discrepancy of the number density of DRGs with $K\lesssim22$ between 
 the HDF-S and HDF-N (e.g., \cite{dic00}, \cite{fon00}). They
reported that only two DRGs (one at $K<21$, one at $21<K<22$) exist 
in the HDF-N, while many DRGs are found in the HDF-S. This can be seen in our  
 data. In Figure \ref{fig:xy}, we plot the spatial distribution of DRGs in
the MOIRCS deep field.
The region of the original HDF-N is also showed, and there are only
two DRGs with $K<22$ and one DRG with $22<K<23$ 
 in this region. On the other hand, there are many DRGs over 
wide range of $K$-band magnitude outside the HDF-N region, and not only
the bright DRGs but also the faint ones with $22<K<23$ appear to show   
a relatively strong clustering. 
Such a spatial distribution of DRGs 
 indicates that the small survey areas of the HDFs could 
introduce large uncertainty and the wide area survey is essential for 
the estimation of the number density of these galaxies.
We will present the detailed analysis 
of the clustering property of
these galaxies in Ichikawa et al. (in preparation).

\begin{figure}
\begin{center}
\FigureFile(85mm,85mm){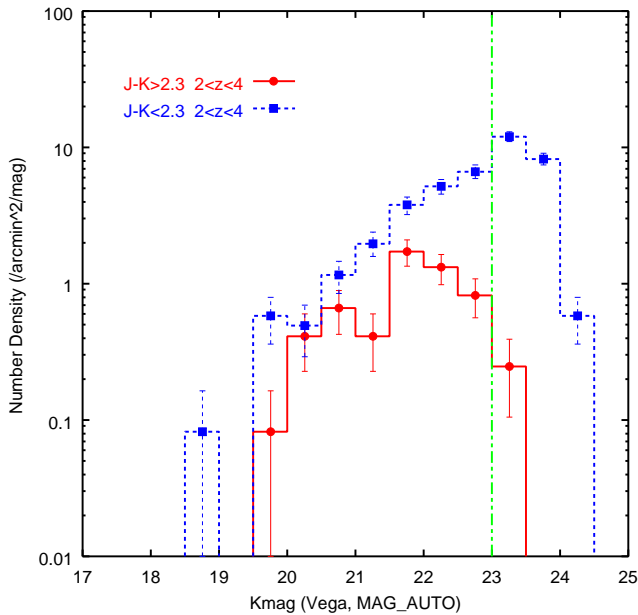}
\end{center}
\caption{$K$-band number counts of galaxies with $2<z_{\rm phot}<4$ in
the MOIRCS deep imaging field.
Circles show those galaxies with $J-K>2.3$ (i.e., DRGs) and squares
represent those with $J-K<2.3$.
Errorbars are based on Poissonian.
The completeness limit is also shown (dotted-dash line). 
}
\label{fig:z3num}
\end{figure}
What does the deficit of the faint DRGs at $K>22$ mean ?
The previous studies found that most DRGs exist at $2<z<4$
(\cite{van03}, \cite{red05}, \cite{for04}).
 Our photometric 
redshift estimation with the HST/ACS and MOIRCS data also suggests 
the similar redshift distribution; 69 out of 86 DRGs with photometric
redshift (those detected in more than two bands) have $2<z_{\rm phot}<4$
(Konishi et al. in preparation). Keeping in mind that some significant
fraction of 
bright DRGs confirmed to be at $z<2$ (e.g., \cite{dad04}, 
\cite{con06}) and that the reliability of photometric redshift 
for the faint DRGs has not yet confirmed by spectroscopy, we assume
that most DRGs in our data lie at $2<z<4$ in the following discussion.

The observed color of $J-K=2.3$ roughly corresponds to the rest-frame 
$U-V\sim0.4$ ($U-B\sim-0.1$) at $2<z<4$, which is similar with those of Sc-Sd
galaxies seen 
in the present universe. 
Such a red color for high-z objects (about 2-3 Gyr age of the
universe at $2<z<4$) is considered to be produced by
Balmer/4000\AA-break of relatively old stellar population 
 and/or heavy dust extinction as mentioned in Section 1.
Therefore, the turnover of the
number counts of DRGs indicates that while there are many bright (massive) 
galaxies with the old population and/or dusty star formation activity, 
the number of the faint galaxies with the similar red color is relatively
small at that epoch. 
Labb{\'e} et al. (2005) investigated the
stellar mass and SED of DRGs and LBGs with $K<22$ and found that DRGs
dominate the high-mass end at $z\sim3$. On the other hand, our result
from the wide and deep data 
could suggest that the low-mass 
end 
is dominated by galaxies with blue colors, which do not 
satisfy the criterion of DRG.
In Figure \ref{fig:z3num}, we show the $K$-band number counts of
galaxies with $2<z_{\rm phot}<4$ for DRGs ($J-K>2.3$) and those with
$J-K<2.3$, separately. It is seen that the fraction of those with
$J-K<2.3$ increases with decreasing $K$-band flux at $K\gtrsim22$, while the
number density of DRGs is comparable to that of the bluer objects  
at $K\sim$20-21. 
Kajisawa and Yamada (2005, 2006) showed that 
most low-mass galaxies have the 
blue rest-frame $U-V$ color and the more massive 
galaxies tend to have the redder color even at $z\sim2.5$. 
Kajisawa \& Yamada (2006) also evaluated that the transition mass between
these red and blue 
populations is about 6$\times10^{10}$M$_{\odot}$ at $z\sim2.5$, 
and in fact most of galaxies with M$_{*}\lesssim10^{10}$M$_{\odot}$ 
 have the rather blue color of $U-V\lesssim0.1$  (their Figure 2).
Since the magnitude of $K=22$ of galaxies with $J-K\sim2.3$
 corresponds to about 1-6$\times10^{10}$M$_{\odot}$ at $2<z<4$ 
(we calculated the stellar mass as in \cite{kaj06}, 
using the GALAXEV population synthesis model of 
\cite{bru03}), 
the deficit of DRGs at
$K>22$ could be explained by such a mass-dependent color distribution, 
if the similar trend continues to $z\sim4$. 

We plan to perform the statistical analysis of the SEDs of these
galaxies over the wide range of luminosity (and mass)  
in order to constrain their physical properties 
in a forthcoming paper. 

\bigskip
We would like to thank the Subaru Telescope staff for their invaluable 
assistance.
We would also like to thank Dr. Masato Onodera for useful discussions
and an anonymous referee for invaluable comments.
This study is based on data collected at Subaru Telescope, which is operated by
the National Astronomical Observatory of Japan. 
Data reduction/analysis was carried out on ``sb'' computer system
operated by the Astronomical Data Analysis Center (ADAC) and Subaru
Telescope of the National Astronomical Observatory of Japan.
The Image Reduction and Analysis Facility (IRAF) is distributed by the 
National Optical Astronomy Observatories, which are operated by the 
Association of Universities for Research in Astronomy, Inc., under
cooperative agreement with the National Science Foundation.
%


\begin{thebibliography}{}
\bibitem[Bershady et al. (1998)]{ber98} Bershady, M.~A., 
Lowenthal, J.~D., \& Koo, D.~C.\ 1998, ApJ, 505, 50 
\bibitem[Bertin \& 
Arnouts(1996)]{ber96} Bertin E., Arnouts S., 1996, A\&AS, 
117, 393 
\bibitem[Bruzual \& 
Charlot(2003)]{bru03} Bruzual G., Charlot S., 2003, MNRAS, 
344, 1000 
\bibitem[Conselice et al.(2006)]{con06} Conselice, C.~J., et 
al.\ 2006, to appear in ApJ, astro-ph/0607242 
\bibitem[Daddi et al.(2004)]{dad04} Daddi, E., Cimatti, A., 
Renzini, A., Fontana, A., Mignoli, M., Pozzetti, L., Tozzi, P., \& 
Zamorani, G.\ 2004, \apj, 617, 746 
\bibitem[Dickinson et 
al.(2000)]{dic00} Dickinson M., et al., 2000, ApJ, 531, 624 
\bibitem[Fontana et 
al.(2000)]{fon00} Fontana A., D'Odorico S., Poli F., 
Giallongo E., Arnouts S., Cristiani S., Moorwood A., Saracco P., 2000, AJ, 
120, 2206 
\bibitem[F{\"o}rster Schreiber et 
al.(2004)]{for04} F{\"o}rster Schreiber N.~M., et al., 2004, 
ApJ, 616, 40
\bibitem[F{\"o}rster Schreiber et 
al.(2006)]{for06} F{\"o}rster Schreiber N.~M., et al., 2006, 
AJ, 131, 1891 
\bibitem[Foucaud et 
al.(2006)]{fou06} Foucaud S., et al., 2006, submitted
to MNRAS, astro-ph/0606386 
\bibitem[Franx et al.(2003)]{fra03} 
Franx M., et al., 2003, ApJ, 587, L79 
\bibitem[Giavalisco et 
al.(2004)]{gia04} Giavalisco M., et al., 2004, ApJ, 600, L93
 \bibitem[Grazian et 
al.(2006)]{gra06} Grazian A., et al., 2006, A\&A, 453, 507 
\bibitem[Ichikawa et al.(2006)]{ich06}
Ichikawa T., et al., 2006, in Proc. of SPIE, Vol. 6269, in press
\bibitem[Iwata et al.(2005)]{iwa05} Iwata I., Inoue A.~K., Burgarella 
D., 2005, A\&A, 440, 881 
\bibitem[Kajisawa \& 
Yamada(2005)]{kaj05} Kajisawa M., Yamada T., 2005, ApJ, 618, 91
\bibitem[Kajisawa \& 
Yamada (2006)]{kaj06} Kajisawa M., Yamada T., 2006, ApJ, 650, 12 
\bibitem[Knudsen et 
al.(2005)]{kun05} Knudsen K.~K., et al., 2005, ApJ, 632, L9
\bibitem[Kriek et al.(2006)]{kri06} 
Kriek M., et al., 2006, ApJ, 645, 44 
\bibitem[Labb{\'e} et 
al.(2003)]{lab03} Labb{\'e} I., et al., 2003, AJ, 125, 1107
\bibitem[Labb{\'e} et 
al.(2005)]{lab05} Labb{\'e} I., et al., 2005, ApJ, 624, L81
 \bibitem[Maihara et 
al.(2001)]{mai01} Maihara T., et al., 2001, PASJ, 53, 25
\bibitem[Minowa et al.(2005)]{min05} 
Minowa Y., et al., 2005, ApJ, 629, 29 
\bibitem[Papovich et 
al.(2006)]{pop06} Papovich C., et al., 2006, ApJ, 640, 92 
\bibitem[Quadri et al.(2006)]{qua06} 
Quadri R., et al., 2006, submitted to ApJ, astro-ph/0606330 
\bibitem[Reddy et al.(2005)]{red05} 
Reddy N.~A., Erb D.~K., Steidel C.~C., Shapley A.~E., Adelberger K.~L., 
Pettini M., 2005, ApJ, 633, 748 
\bibitem[van Dokkum et al.(2003)]{van03}
van Dokkum P.~G., et al., 2003, ApJ, 587, L83 
\bibitem[van Dokkum et 
al. (2004)]{van04} van Dokkum P.~G., et al., 2004, ApJ, 611, 
703 
\bibitem[Williams et 
al. (1996)]{wil96} Williams R.~E., et al., 1996, AJ, 112, 
1335
\end{thebibliography}
\end{document}